\pdfoutput=1
\documentclass[10pt,journal,compsoc]{IEEEtran}

\ifCLASSOPTIONcompsoc
  \usepackage[nocompress]{cite}
\else
  \usepackage{cite}
\fi
\usepackage{wasysym}
\usepackage{hyperref}
\usepackage{tabularx}
\usepackage{graphicx}
 \usepackage{amssymb}
 \usepackage{multirow}
  \usepackage{wrapfig}
\usepackage{color}
\usepackage{enumitem}
\usepackage{dblfloatfix} 
\usepackage[table]{xcolor}
\usepackage{caption}

\newcommand{\etal}{et al. }
\newcommand{\IT}[1]{{\bf%
DODGE(\ifx*#1$\mathcal{E}$\else#1\fi)}}

\begin{document}
\title{How  to ``DODGE''      Complex  Software Analytics}

\author{Amritanshu~Agrawal, 
        Wei~Fu,
        Di~Chen, 
        Xipeng Shen~\IEEEmembership{IEEE Senior Member}, Tim~Menzies,~\IEEEmembership{IEEE Fellow}\thanks{A. Agrawal (aagrawa8@ncsu.edu) works at Wayfair.
W. Fu (wfu@ncsu.edu) works at   Landing.AI.
D. Chen (dchen20@ncsu.edu) works at Facebook. 
X. Shen (xshen5@ncsu.edu), T. Menzies (timm@ieee.org) are   NC State professors.}}

\markboth{IEEE Transactions in Software Engineering, ~Vol.~XXX, No.~XX, August~XXXX}%
{Agrawal \MakeLowercase{\textit{\etal}}: How  to ``DODGE''      Complex  Software Analytics?}

\IEEEtitleabstractindextext{%
\begin{abstract} 
Machine learning techniques applied to
software engineering tasks can be improved by hyperparameter optimization,
i.e., automatic tools that find good settings for a learner's control parameters.
We show that such hyperparameter optimization can be 
unnecessarily slow, particularly 
when the optimizers
waste time exploring  
``redundant  tunings'', i.e.,
 pairs of tunings which lead to  indistinguishable    results.
By ignoring redundant tunings, 
 {\IT*}, a  tuning tool,
  runs orders of magnitude faster, while also
generating
  learners with more accurate predictions  than  
  seen in
  prior state-of-the-art approaches.  
\end{abstract}
}

\maketitle

\IEEEdisplaynontitleabstractindextext

\IEEEpeerreviewmaketitle

\newcommand{\TEXT}{black}

\newenvironment{redish}{\color{\TEXT}}

\newcommand{\tion}[1]{\S\ref{sect:#1}}
\newcommand{\eq}[1]{Equation~\ref{eq:#1}}
\newcommand{\fig}[1]{Figure~\ref{fig:#1}}

\newcommand{\tbl}[1]{Table~\ref{tbl:#1}}
\newcommand{\bi}{\begin{itemize}[leftmargin=0.4cm]}
	\newcommand{\ei}{\end{itemize}}
\newcommand{\be}{\begin{enumerate}[leftmargin=0.4cm]}
	\newcommand{\ee}{\end{enumerate}}

\ifCLASSOPTIONcompsoc
\IEEEraisesectionheading{\section{Introduction}\label{sect:intro}}
\else
\section{Introduction}
\label{sect:intro}
\fi

Fisher et al.~\cite{interactions-with-big-data-analytics} defines software analytics as  a workflow that distills large quantities of low-value
data into  smaller sets of higher value data.  
Such analytics aims at generating insights and building predictors for  software systems. 

 Due to the complexities and computational cost of SE
analytics, Fisher et al warn that    ``the luxuries of interactivity, direct manipulation, and fast system response are gone''.
In fact, they characterize modern cloud-based analytics as a throwback to the 1960s- batch processing mainframes where
jobs are submitted and then analysts wait long for results with ``little insight into what’s really going
on behind the scenes, how long it will take, or how much it’s going to cost''.
Fisher et al. document
 issues seen by  industrial data scientists, one who says
{\em ``Fast iteration is key, but incompatible
with jobs ...  in the cloud. It’s frustrating to wait for hours, only to realize
you need a slight tweak...''}

One impediment to  fast iterations are 
 {\em hyperparameter optimizers}
 that automatically tune  control options for data mining.
 Off-the-shelf
 learners   come
with   defaults for  control parameters, which may be sub-optimal. For example,
in the  distance function  $d(x,y,p)=\left(\sum_i (x_i-y_i)^p\right)^{1/p}$,
a standard default is $p=2$. 
Yet Agrawal et al.~\cite{agrawal2018better}  found that
$p>2$ worked much better for their processing. 

Hyperparameter optimizers   automatically     find better control parameters
 by experimenting with adjustments to the control parameters of a learner~\cite{biedenkapp2018hyperparameter}~\cite{franceschi2017forward}. 
When done using 21st century optimizers (e.g., NSGA-2~\cite{deb00afast}, IBEA~\cite{Zitzler04indicator-basedselection}, MOEA/D~\cite{zhang07}, FLASH~\cite{nair18}), it is now possible to optimize for multiple goals (even when they are competing).
\tbl{options} lists some
tuning options for data pre-processing and machine learning for   two
well-studied 
SE tasks:
\bi
\item

{\em Software defect prediction}
(classifying  modules into ``buggy'' or otherwise~\cite{agrawal2018better,chen2018applications,fu2016tuning,tantithamthavorn2016automated,8263202,menzies2007data,ghotra2015revisiting});
\item
 {\em Software bug report text mining} (to find        severity~\cite{agrawal2018better,oliveira2010ga}).
\ei
\textcolor{\TEXT}{ \tbl{options} is  a partial
list of some of the  tunings that might be explored. 
Even this incomplete sample includes billions of configuration options.}
With enough CPU, automatic hyperparameter  optimizers
can prune those options to
find tunings that  improve the performance
 of software quality predictors~\cite{agrawal2018better,fu2016tuning,liu2010evolutionary,sarro2012further,8263202,tantithamthavorn2016automated,zhong2004,treude2018per,oliveira2010ga}. 
For example, Tantithamthavorn et al.~\cite{tantithamthavorn2016automated,8263202} showed that 
tuning can convert     bad learners into very good ones.

The problem  with hyperparameter optimization is finding enough CPU. 
The  cost of running a data miner through  all those  options is very high, requiring days to weeks to decades of CPU~\cite{abs-1807-11112,8263202,tantithamthavorn2016automated,wang2013searching,treude2018per,xia18}.
For many years, we have addressed these long CPU times  via cloud-based CPU farms.
Fisher et al.~\cite{interactions-with-big-data-analytics} warn that cloud computation is a heavily monetized environment that charges for all their services (storage, uploads, downloads, and CPU time). While each small part of that service is cheap, the total annual cost to an organization can be exorbitant.

 \begin{table*}
\footnotesize
\captionsetup{font=footnotesize}
\caption{Hyperparameter tuning options explored in this paper.
Options in learners from  recent SE papers on hyperparameter optimization~\cite{ghotra2015revisiting,fu2016tuning,agrawal2018better,agrawal2018wrong} then consulting the documentation of a widely-used data mining library
(Scikit-learn~\cite{pedregosa2011scikit}). Randint,  randuniform and randchoice are all random functions to choose either integer, float, or a choice among the parameter ranges.}\label{tbl:options}
\begin{tabular}{|p{.4\linewidth}p{.56\linewidth}|}\hline
\multicolumn{2}{|p{.96\linewidth}|}{\textbf{DATA PRE-PROCESSING}} \\
\multicolumn{2}{|p{.96\linewidth}|}{~} \\

\noindent
Software defect prediction: \vspace{3mm}

\bi 
\item StandardScaler
\item MinMaxScaler
\item MaxAbsScaler
\item RobustScaler(quantile\_range=(a, b))
    \bi
    \item a,b = randint(0,50), randint(51,100)
    \ei
\item KernelCenterer
\item QuantileTransformer(n\_quantiles=a,\newline output\_distribution=c, subsample=b)
    \bi
    \item a, b = randint(100, 1000), randint(1000, 1e5)
    \item c = randchoice([`normal',`uniform'])
    \ei
\item Normalizer(norm=a)
    \bi
    \item a = randchoice([`l1', `l2',`max'])
    \ei
\item Binarizer(threshold=a)
    \bi
    \item a = randuniform(0,100)
    \ei
    \item SMOTE(a=n\_neighbors, b=n\_synthetics,\newline c=Minkowski\_exponent)
    \bi
    \item a,b = randint(1,20),randchoice(50,100,200,400)
    \item c = randuniform(0.1,5)
    \ei
\ei

&
Text mining: \vspace{3mm}
 
\bi
\item CountVectorizer(max\_df=a, min\_df=b)
    \bi
    \item a, b = randint(100, 1000), randint(1, 10)
    \ei
\item TfidfVectorizer(max\_df=a, min\_df=b, norm=c)
    \bi
    \item a, b,c = randint(100, 1000), randint(1, 10), randchoice([`l1', `l2', None])
    \ei
\item HashingVectorizer(n\_features=a, norm=b)
    \bi
    \item a = randchoice([1000, 2000, 4000, 6000, 8000, 10000])
    \item b = randchoice([`l1', `l2', None])
    \ei
\item LatentDirichletAllocation(n\_components=a, doc\_topic\_prior=b,\newline topic\_word\_prior=c,
                                   learning\_decay=d,
                                   learning\_offset=e,batch\_size=f)
    \bi
    \item a, b, c = randint(10, 50), randuniform(0, 1),  randuniform(0, 1)
    \item d, e  = randuniform(0.51, 1.0), randuniform(1, 50), 
    \item f = randchoice([150,180,210,250,300])
    \ei
\ei
\\\hline
\multicolumn{2}{|p{.96\linewidth}|}{\textbf{LEARNERS}}\\
\multicolumn{2}{|p{.96\linewidth}|}{~} \\
 \multicolumn{2}{|p{.96\linewidth}|}{

\noindent
Software defect prediction and text  mining:\vspace{3mm}
\begin{itemize}[noitemsep,topsep=0pt]
\item DecisionTreeClassifier(criterion=b, splitter=c, min\_samples\_split=a)
    \bi
   \item a, b, c= randuniform(0.0,1.0), randchoice([`gini',`entropy']),  randchoice([`best',`random'])
   \ei
\item RandomForestClassifier(n\_estimators=a,criterion=b,  min\_samples\_split=c)
    \bi
   \item a,b,c = randint(50, 150), randchoice(['gini', 'entropy']), randuniform(0.0, 1.0) 
   \ei
\item LogisticRegression(penalty=a, tol=b, C=float(c))
    \bi
    \item a,b,c=randchoice([`l1',`l2']), randuniform(0.0,0.1), randint(1,500)
    \ei 
\item MultinomialNB(alpha=a)
    \bi
    \item a= randuniform(0.0,0.1)
    \ei 
\item KNeighborsClassifier(n\_neighbors=a, weights=b, p=d, metric=c)
    \bi
    \item a, b,c  = randint(2, 25), randchoice([`uniform', `distance']), randchoice([`minkowski',`chebyshev'])
    \item if c=='minkowski': d= randint(1,15)  else:  d=2
    \ei

\end{itemize}
}
\\\hline
\end{tabular}

\end{table*}
 


Recently it was discovered how to (a)~save most of that CPU cost while at the same time
\begin{wrapfigure}{r}{1.3in}
~~~~\includegraphics[width=1in]{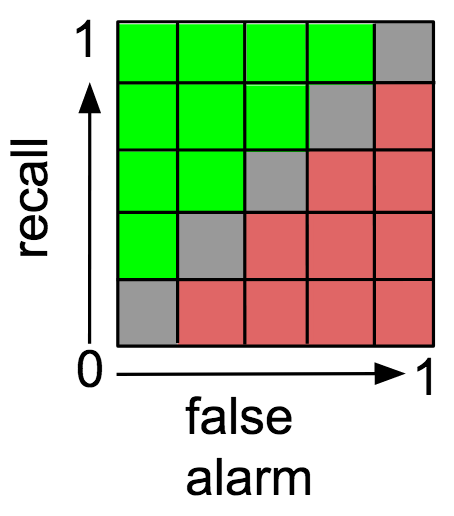}
\caption{For \ \mbox{$\mathcal{E}=0.2$},
  outputs  have      25 cells. Green
  cells are preferred (high recall 
  and low false alarms). }\label{fig:grid}
\end{wrapfigure}
 (b)~find better tunings.
\textcolor{\TEXT}{As   discussed later,
 a method called  ``FFtrees''~\cite{phillips2017FFtrees}  (which
just selects a best model within   a small
forest of shallow decision trees)
generates much better predictions than  supposed state-of-the-art results obtained
after CPU intensive  tuning~\cite{chen2018applications}.}  This is    strange   
  since  standard  tuning tries  thousands of options, but FFtrees tries just a dozen.

\textcolor{\TEXT}{To explain these FFtree results~\cite{chen2018applications}, we observe
that (a)~a learner assessed by $p$ performance scores has a $p$ dimensional output space;
and (b)~there is  some variation $\mathcal{E}$  where a learner's performance appears in that space.}
As shown  in \fig{grid}, if
\mbox{$\mathcal{E}=0.2$} then the 2 performance scores ($p=2$) output space divides into 
\textcolor{\TEXT}{$1/\mathcal{E}^p=1/{0.2}^2=25$ cells}.
That is, 
if we explored more than 25 tunings, certain
pairs of tunings  would be {\em redundant} (i.e.,  would have  very similar outcomes).

It turns out there are better ways to avoid redundant tunings than FFtrees.
Our method {\IT*} learns  to ignore redundant tunings (parameter settings including which classifier and preprocessor to use)
those that  fall within $\mathcal{E}$ of other results.
When tested  on defect 
prediction and text mining,
{\IT*} terminated after  fewer evaluations than standard optimizers.
Also, it produced better performance scores than state-of-the art research articles (for the two well-studied SE tasks listed before~\cite{fu2016tuning,agrawal2018better,chen2018applications,agrawal2018wrong,Panichella:2013,ghotra2015revisiting}).
We conjecture that other methods   perform relatively worse  since they do not appreciate just how small  
the output space is. Hence, those other methods waste  CPU as they
 struggle  to cover  billions of redundant tuning options like \tbl{options} (most of which yield  indistinguishable   results).

This article introduces and evaluates {\IT*}.    \tion{fft}
describes   how FFtrees lead to
 the design of {\IT*} (in \tion{dodge}).
 \tion{rq} then answers     the following research questions.
 
{\bf RQ1: Is {\IT*} too complicated?  How to find appropriate value of $\mathcal{E}$?}
We can not recommend a  method
if it is too complex to use.
Fortunately, we    show that  it is easy to find {\IT*}'s  parameters since
its   success is 
not altered by large  
changes to $\mathcal{E}$.
 
\textcolor{\TEXT}{{\bf RQ2: How does {\IT*} compare to recent prominent
defect prediction and hyperparameter optimization results?}
When compared to recent tuning papers at  IST'16, ICSE'18 and FSE'18 results~\cite{fu2016tuning,ghotra2015revisiting,chen2018applications}, 
{\IT*}  explored a much larger parameter search space and exhibited much faster termination.
Also, in terms of goal performance:
\bi
\item
{\IT*}   out-performed
an ICSE'15 article exploring different learners for
defect prediction~\cite{ghotra2015revisiting} by around 50\% and 40\%  for  {\em d2h} and {\em Popt(20)} respectively\footnote{
{\em d2h} scores highest    for models with high recalls and low false alarms while 
{\em Popt(20)} scores highest when many defects are localized to a small part of the code. 
For full details on these measures, and why we use them, see    \tion{goals}.}.
\item
 {\IT*}  also did better than the IST'16 journal article that demonstrated
the value of  tuning
for learners~\cite{fu2016tuning} by about 30\% and 10\%  (for {\em d2h} and {\em Popt(20)}).
\item
This approach also does better than
the ICSE'18 article that advocated  to   tune  data pre-preprocessors~\cite{agrawal2018better} by about 10\% and 5\% on an average (for {\em d2h} and {\em Popt(20)}).
\item
Further, {\IT*} also does better by   10\% and 5\% (for {\em d2h} and {\em Popt(20)}) than
the FSE'18 article mentioned earlier 
that reported FFtrees~\cite{chen2018applications}. 
\ei
}

{\bf RQ3: Is {\IT*} only useful for defect prediction?}
In order to stress test our methods,  we must apply {\IT*}
to some harder task than defect prediction. 
Software bug report text mining  is a
harder task than defect prediction since
the
latter only process a few dozen attributes while former
task have tens of thousands of attributes. 
\textcolor{\TEXT}{For   text mining, we show 
{\IT*} performs better than
the IST'18 journal article that  showed the value 
of tuning for SE text mining applications~\cite{agrawal2018wrong}  by about 20\% on an average for {\em d2h}.
Also {\IT*}
performs better than
the ICSE'13 article that applied genetic algorithms to
learn the settings for a text miner~\cite{Panichella:2013} by about 20\% on an average for {\em d2h}.
As with the defect prediction studies,
for both these IST'18 and ICSE'13 papers, 
{\IT*}  explored a much larger parameter search space and exhibited much faster termination.}

 
 \begin{figure}[!t]
\centering
\includegraphics[width=\linewidth]{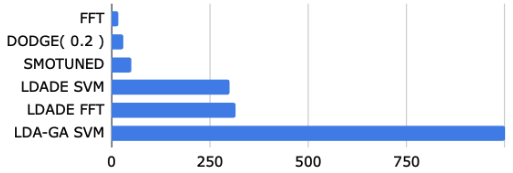}
\captionsetup{font=footnotesize}
\caption{\textcolor{\TEXT}{Comparisons of the computational cost of the different methods studied in the paper.  Here, the  computational cost  is
measured in terms of the number of evaluations required to find a model. The y-axis shows various methods, discussed later in this paper.
The essential point of this figure is that methods that know how
to avoid redundant tunings (i.e. FFT and {\IT*}) cost orders of magnitude
less than otherwise. }
}\label{fig:timings}
\end{figure}

\textcolor{\TEXT}{From our findings, we could recommend FFtrees if the goal
is only to produce succinct, approximate summaries of the factors
that matter in the data.  Also, as shown in  \fig{timings}, FFtrees are slightly faster than
{\IT*}.}

\textcolor{\TEXT}{That said, if the goal is maximizing predictive prowess 
then  we must caution that   FFtree's faster generation of smaller models comes at a price-
FFtrees  usually generates significantly weaker predictions than   {\IT*} (see the RQ2 and RQ3 results, discussed later).}
Another reason to recommend {\IT*} is that it
 generates better  predictors than numerous  recent SE state-of-the-art research articles~\cite{fu2016tuning,agrawal2018better,chen2018applications,agrawal2018wrong,Panichella:2013,ghotra2015revisiting}. 
 

But  more fundamentally, the other reason to explore {\IT*} is that it tests the theory that much better hyperparameter optimizers can be built by assuming the  output space divides into just
a few regions of size 
  $\mathcal{E}$. {\IT*} is one way to exploit this effect.
  We believe that  further research could be performed in many others ways
  (e.g., different learners, better visualizations and/or explanations of analytics, faster implementations of other tools).


 \subsection{Relation to Prior Work}\label{sect:prior}
 
\textcolor{\TEXT}{All the  {\IT*} work is novel for this paper (this research team invented {\IT*} and this is the first
publication to discuss it). As to the work on FFtrees, for defect prediction, this paper includes
the prior results with that of Chen et al. results~\cite{chen2018applications}. The application of FFtrees to text mining (in this paper) is
a novel result.}

\section{Background}\label{sect:background}
This section describes the background on defect prediction, and text mining and the corresponding data and methods which are considered baselines.
 
\subsection{Text Mining}
\label{sect:tm}

\textcolor{\TEXT}{
Many SE project artifacts come in the form of {\em unstructured text} such as
word processing files, slide presentations, comments, 
Github issue reports, etc.
According to White~\cite{white05}, 80\% of business is conducted on unstructured data, 85\%  of all data stored is held in an unstructured format and unstructured data doubles every three months. 
Nadkarni and Yezhkova~\cite{nadkarni2014structured}
say that 1,600 Exabytes of data appears
in unstructured sources  and that
each year, humans generate more unstructured artifacts than structured.}

\textcolor{\TEXT}{Lately, there have been   much interest in SE text mining~\cite{menzies2008improving,menzies2008automated,Panichella:2013,agrawal2018wrong,xu2016predicting,majumder18}
since it covers a much wider range of SE activities. 
Text mining is  harder   than other case studies (like defect prediction) due to presence of free form natural language  which is semantically very complex and may not conform to any known grammar. 
In practice,   text documents   require   tens of thousands of attributes (one for each word).
For example, consider  NASA's software
project and issue tracking systems (or PITS)~\cite{menzies2008improving, menzies2008automated} that contain text
discussing
bugs and changes in source code. As shown in \tbl{data_text}, our text data contains tens to hundreds of thousands
of words (even when reduced to unique words, there are still 10,000+ unique words).
}

 \begin{table}
\begin{center}

\caption{Dataset statistics. Data comes from the SEACRAFT repository: \url{http://tiny.cc/seacraft}}
\label{tbl:data_text}
\begin{tabular}{c@{~}|r@{~}|r@{~}|r@{~}}
\begin{tabular}[c]{@{}c@{}} \textbf{Dataset} \end{tabular} & \begin{tabular}[c]{@{}c@{}} \textbf{No. of Documents}\end{tabular} & \textbf{No. of Unique Words} & \begin{tabular}[c]{@{}c@{}} \textbf{Severe \%}\end{tabular} \\ \hline
PitsA & 965 & 155,165  & 39  \\ 
PitsB &   1650 & 104,052  & 40  \\  
PitsC &   323 & 23,799  & 56 \\ 
PitsD &   182 & 15,517 & 92  \\  
PitsE & 825 & 93,750  & 63 \\ 
PitsF & 744 & 28,620 & 64 \\ 
\end{tabular}
\end{center} 
\end{table}

\subsubsection{Data and Algorithms for Text Mining}

\textcolor{\TEXT}{\tbl{data_text} describes our 
 PITS data, which comes from   six different NASA systems (which we label PitsA, PitsB,...etc). 
 For this study, all  datasets were preprocessed using the usual text mining filters~\cite{feldman2006j}.
We implemented   stop words removal using NLTK toolkit~\cite{bird2006nltk} 
(to ignore very common short words such as  ``and'' or ``the'').
Next, 
  Porter's stemming filter~\cite{Porter1980} was used  to delete uninformative word endings
  (e.g., after performing stemming, all the following words would be rewritten
  to ``connect'': ``connection'', ``connections'',
``connective'',          
``connected'',
  ``connecting'').
After that,  {\IT*} selected other pre-processors using  the space of options from  \tbl{options}. }

\textcolor{\TEXT}{A standard    text mining learner is SVM (support vector machine).
A drawback with SVM is that its models may not be human comprehensible.
Finding insights among unstructured text
is  difficult unless we can search, characterize, and classify the
textual data in a meaningful way. One of the common techniques for
finding related topics within unstructured text (an area called
topic modeling) is the Latent Dirichlet allocation (LDA)~\cite{blei2003latent}.
LDA clusters text into ``topics'' defined by the high-frequency words in that cluster.
For example, the topics found by LDA for one of our PITS data sets are shown in \tbl{topics}.
We studied LDA since it is a widely-used technique  in  prominent  SE research articles~\cite{agrawal2018wrong}.}

 \begin{table}[!t]
\begin{center}
\caption{Top 10 topics found by LDA for PitsA dataset from\tbl{data_text}. Within 
each topic, the weight of words decreases exponentially  left to right across the order shown here.
The words here are truncated (e.g., ``software'' becomes ``softwar'') due to stemming.}
\label{tbl:topics}
 \resizebox{\linewidth}{!}{
\begin{tabular}{r@{=~}l}
Topics & Top words in topic\\\hline
  01 & command engcntrl section spacecraft unit icd tabl point referenc indic \\
  02 & softwar command test flight srobc srup memori script telemetri link \\
  03 & file variabl line defin messag code macro initi use redund \\
  04 & file includ section obc issu fsw code number matrix src \\
  05 & mode safe control state error power attitud obc reset boot \\
  06 & function eeprom send non uplink srup control load chang support \\
  07 & valu function cmd return list ptr curr tss line code \\
  08 & tabl command valu data tlm load rang line count type \\
  09 & flight sequenc link capabl spacecraft softwar provid time srvml trace \\
  10 & line messag locat column access symbol file referenc code bld
\end{tabular}
}
\end{center}
\end{table}

 \begin{table}[!t]
\begin{center}

\caption{Document Topic distribution found by LDA for PitsA dataset}
\label{tbl:features}
\begin{tabular}{r|r@{~~~}r@{~~~}r@{~~~}r@{~~~}r@{~~~}r@{~~~}r@{~~~}r@{~~~}r@{~~~}r|c}
Issue & \multicolumn{10}{c|}{10 Topics} & Severe? \\\hline
  01 &  .60 &  .10 &.00& .15 & .00& .05 & .03 & .04 & .03 &.00& y \\
  02 & .10 & .03 & .02 &.00& .03 & .02 & .15 & .65 &.00&.00& n  \\
  03 &.00& .20 & .05 & .05 &.00& .60 & .02 & .03 & .03 & .02 & n  \\
  04 & .03 & .01 & .01 & .10 & .15 &.00& .70 &.00&.00&.00& y  \\
  etc &
\end{tabular}
\end{center}
\end{table}

\textcolor{\TEXT}{LDA is controlled by various parameters (see \tbl{options}).
At ICSE'13,
Panichella et al.~\cite{Panichella:2013} used a genetic algorithm  to tune  their LDA
text miners.
More recently, in the IST'18 journal article,
Agrawal et al.~\cite{agrawal2018wrong} saw that  differential evolution  can out-perform genetic algorithms
for tuning LDA.}

\textcolor{\TEXT}{A standard pre-processor for text mining is {\em vectorization}; i.e., replace the raw observations
of wordX appearing in documentY with some more informative statistic.
For example,  Agrawal et al. converted the PITS text data  into the vectors
of \tbl{features}. The cells in that  table shows how much
each issue report matches each topic (and the final column shows  the issue severity of that report).
\tbl{options} lists the options for the LDA vectorization, plus three other vectorization methods.}

\subsection{Defect Prediction}\label{sect:dp}

\textcolor{\TEXT}{Software developers are smart, but sometimes make mistakes. Hence, it is essential to test software before the deployment ~\cite{orso2014software,barr2015oracle,yoo2012regression, myers2011art}. 
 \textcolor{\TEXT}{Software quality assurance budgets are finite but increasing assessment effectiveness by some
 linear amount can take   exponentially more effort}~\cite{fu2016tuning}. Therefore,  standard practice is to apply the best available methods on code sections that seem most critical and bug-prone.
Software bugs are not evenly distributed across the project~\cite{hamill2009common,koru2009investigation, ostrand2004bugs,misirli2011ai}.  Hence, 
a useful way to perform software testing is to allocate most assessment budgets to
the more defect-prone parts in software projects.   
Software defect predictors are never 100\% correct. But
they can be used to suggest where to focus more expensive methods. }

\textcolor{\TEXT}{There is  much  commercial interest  in  defect prediction.  In a survey of  395 practitioners from 33 countries and five continents,
 Wan et al.~\cite{wan18} found that over 90\% of
 the respondents were willing to adopt defect prediction
 techniques. 
When Misirli et al.~\cite{misirli2011ai} built a  defect prediction model  for a telecommunications company, those  models
could  predict 87\% of files with  defects.
Those models also decreased inspection efforts by 72\%, and hence reduced post-release defects by 44\%. }

Software defect predictors not only save labor compared with traditional manual methods, but they are also competitive with certain automatic methods.  
A recent study at ICSE'14, Rahman et al. ~\cite{rahman2014comparing} compared (a) static code analysis tools FindBugs, Jlint, and PMD and (b) static code defect predictors (which they called ``statistical defect prediction'') built using logistic  regression.
They found no significant differences in the cost-effectiveness of these approaches.  

Given this equivalence, it is significant to
note that static code defect prediction can be quickly adapted to new languages by building lightweight parsers to extract static code metrics such as \tbl{ck}. The same is not true for static code analyzers - these need extensive modification before they can be used in new languages.


 \begin{table}[!b]
\footnotesize
\caption{OO code metrics used for the defect prediction studies of this article. \textcolor{\TEXT}{For definitions on code metrics, please refer table 1 of \cite{agrawal2018better}.}
	   Last line, shown in \colorbox{lightgray}{gray}, denotes the dependent variable.}\label{tbl:ck}
\begin{center}

			\begin{tabular}{c|l}\hline
				amc & average method complexity \\\hline
				avg\, cc & average McCabe \\\hline
				ca & afferent couplings \\\hline
				cam & cohesion among classes \\\hline
				cbm & coupling between methods \\\hline
				cbo & coupling between objects \\\hline
				ce & efferent couplings \\\hline
				dam & data access\\\hline
				dit & depth of inheritance tree\\\hline
				ic & inheritance coupling\\\hline
				lcom (lcom3) & 2 measures of lack of cohesion in methods \\\hline
				loc & lines of code \\\hline
				max\, cc & maximum McCabe\\\hline
				mfa & functional abstraction\\\hline
				moa &  aggregation\\\hline
				noc &  number of children\\\hline
				npm & number of public methods\\\hline
				rfc & response for a class\\\hline
				wmc & weighted methods per class\\\hline
				\rowcolor{lightgray}
				defects & Boolean: where defects found in bug-tracking\\
			\end{tabular}
		
	\end{center}
\end{table}

\subsubsection{Data and Algorithms for Defect Prediction}\label{sect:daa}

Our defect predictors where applied to the data described in
\tbl{versions}.
As shown in \tbl{versions},
this data is available for multiple software versions  (from  http://tiny.cc/seacraft). 
This is important since, an important principle of data mining is not to test on the data used in training. 
There are many ways to design a experiment that satisfies this principle.
Some of the methods where we do not test data mining on training data itself have limitations too; e.g., leave-one-out is too slow for
large data sets and cross-validation mixes up older and newer data~
(such that data from the past may be used to test on future data). In
this work, for each project data, we set the latest version of project data
as the testing data and all the older data as the training data.
For example, we use $\mathit{poi1.5, poi2.0, poi2.5}$ data for training predictors,
and the newer data, $\mathit{ poi3.0}$ is left for testing.

\tbl{versions} illustrates the variability of SE data. 
The data  can be observed to have
imbalanced class frequencies. 
If the target class is not common (as in the camel, ivy, etc test data in \tbl{versions}), it is difficult  to generate a model that can locate it.
A standard trick for  class imbalance is   SMOTE~\cite{Chawla:2002} that  synthetically
create members of the minority class. 
\tbl{options} show controlled parameters of SMOTE.

\begin{table}
\centering
\caption{Statistics of the studied data sets.
\textcolor{\TEXT}{For the training data, the reported statistics
come from the combination of all the versions
used in training.
  In this table,
  the  defective  ratio  represents the  combination of  total  defective ratio  after combining all the software versions used for training (hence,  we only report  one ratio.
} }
\label{tbl:versions}
\resizebox{!}{0.21\linewidth}{
\begin{tabular}{cllll}
\hline
\rowcolor{lightgray} \multicolumn{1}{c|}{}& \multicolumn{2}{c|}{Training Data}& \multicolumn{2}{c}{Testing Data}\\ \cline{2-5} 
\rowcolor{lightgray}\multicolumn{1}{c|}{\multirow{-2}{*}{Project}} & \multicolumn{1}{c|}{Versions} & \multicolumn{1}{c|}{\% of Defects} & \multicolumn{1}{c|}{Versions} & \multicolumn{1}{l}{\% of Defects} \\ 
\hline

\multicolumn{1}{l}{Poi}& \multicolumn{1}{l}{1.5, 2.0, 2.5}& \multicolumn{1}{l}{426/936 = 46\%}& \multicolumn{1}{c}{3.0}& \multicolumn{1}{l}{281/442 = 64\%} \\ \hline
\multicolumn{1}{l}{Lucene}& \multicolumn{1}{l}{2.0, 2.2}& \multicolumn{1}{l}{235/442 = 53\%}& \multicolumn{1}{c}{2.4}& \multicolumn{1}{l}{ 203/340 = 60\%} \\ \hline
\multicolumn{1}{l}{Camel}& \multicolumn{1}{l}{1.0, 1.2, 1.4}& \multicolumn{1}{l}{374/1819 = 21\%}& \multicolumn{1}{c}{1.6}& \multicolumn{1}{l}{ 188/965 = 19\%
} \\ \hline
\multicolumn{1}{l}{Log4j}& \multicolumn{1}{l}{1.0, 1.1}& \multicolumn{1}{l}{71/244 = 29\%}& \multicolumn{1}{c}{1.2}& \multicolumn{1}{l}{189/205 = 92\%} \\ \hline
\multicolumn{1}{l}{Xerces}& \multicolumn{1}{l}{1.2, 1.3}& \multicolumn{1}{l}{140/893 = 16\%}& \multicolumn{1}{c}{1.4}& \multicolumn{1}{l}{437/588 = 74\%} \\ \hline
\multicolumn{1}{l}{Velocity}& \multicolumn{1}{l}{1.4, 1.5}& \multicolumn{1}{l}{289/410 = 70\%}& \multicolumn{1}{c}{1.6}& \multicolumn{1}{l}{78/229 = 34\%} \\ \hline
\multicolumn{1}{l}{Xalan}& \multicolumn{1}{l}{2.4, 2.5, 2.6}& \multicolumn{1}{l}{908/2411 = 38\%}& \multicolumn{1}{c}{2.7}& \multicolumn{1}{l}{898/909 = 99\%} \\ \hline
\multicolumn{1}{l}{Ivy}& \multicolumn{1}{l}{1.1, 1.4}& \multicolumn{1}{l}{79/352 = 22\%}& \multicolumn{1}{c}{2.0}& \multicolumn{1}{l}{40/352 = 11\%} \\ \hline
\multicolumn{1}{l}{Synapse}& \multicolumn{1}{l}{1.0, 1.1}& \multicolumn{1}{l}{76/379 = 20\%}& \multicolumn{1}{c}{1.2}& \multicolumn{1}{l}{ 86/256 = 34\%} \\ \hline
\multicolumn{1}{l}{Jedit}& \multicolumn{1}{l}{ \begin{tabular}[c]{@{}c@{}}3.2,4.0, 4.1,4.2 \end{tabular}}& \multicolumn{1}{c}{292/1257 = 23\%}& \multicolumn{1}{c}{4.3}& \multicolumn{1}{l}{11/492 = 2\%} \\ \hline
\end{tabular}
}
\end{table}

As to machine learning algorithms, there are many and varied.
At ICSE'15, Ghotra et al.~\cite{ghotra2015revisiting} applied 32 different machine learning algorithms to defect prediction.
In a result consistent with the theme of this article, they found that those 32 algorithms formed into four groups
of Table 9 in \cite{ghotra2015revisiting}
(and the performance of two learners in any one group
were statistically indistinguishable  from each other).




\subsection{Evaluation}\label{sect:goals}
\subsubsection{Measures of Performance}\label{sect:easures}
We eschew   precision and accuracy since these can be inaccurate for data sets where the target class is rare (which is common
in defect prediction data sets)~\cite{Menzies:2007prec}.  
\textcolor{\TEXT}{For 
example, consider a test data set with 20\% defective examples.
A learner could be 80\% accurate for that data set, while still
missing 100\% of  the defective examples.  As to why we deprecate precision,
we refer the interested reader to a prior work~\cite{Menzies:2007prec}.}

Instead, we will evaluate our predictors on metrics
that aggregate multiple metrics.
{\em D2h}, or  ``distance to heaven'', shows how close   scores fall to   ``heaven'' (where recall=1 and false alarms (FPR)=0)~\cite{chen2018applications}. {\em D2h} was used to evaluate both defect predictors as well as text mining.

{\footnotesize\begin{eqnarray} \label{eq:recall}
    \mathit{Recall} & = & \frac{\mathit{True Positives}}{\mathit{True Positives + False Negatives}} \\
    \mathit{FPR} & = & \frac{\mathit{False Positives}}{\mathit{False Positives + True Negatives}} \\
    \mathit{d2h} & = & \frac{ \sqrt{  (1-\mathit{Recall})^2 +   (0-\mathit{FPR   })^2}}{   \sqrt{2}}\label{eq:d2h}                 
\end{eqnarray}}
{\noindent}Here, the  $\sqrt{2}$ term normalizes {\em d2h} to the range zero to one.

For defect prediction, {\em Popt(20)} comments on the effort required {\em after} a defect predictor triggers and humans have to read code, looking
for errors. {\em Popt(20)} is a specialized metric which can be used only with defect predictor.
$Popt(20)=1- \Delta_{opt}$, where $\Delta_{opt}$ is the area
between the effort~(code-churn-based) cumulative lift charts of the optimal
learner and the proposed learner. To calculate {\em Popt(20)}, we divide all the code modules into those predicted to be defective ($D$) or not ($N$). Both sets are then sorted in ascending order of lines of code. The two sorted sets are 
then laid out across the 
 x-axis,
with  $D$ before $N$. This layout means that 
  the x-axis extends from 0 to 100\% where lower values of $x$ are predicted to be more defective than $x$ higher values.
 On such a chart, the y-axis shows what percent of the defects would be recalled if we traverse the code sorted that x-axis order. 
 Following from the recommendations of Ostrand et al.~\cite{ostrand2004bugs},  {\em Popt} is reported at the 20\% point; show how many bugs are find if we  inspect a small portion of the code (20\%).
 
Kamei, Yang et al. ~\cite{yang2016effort,kamei2013large,monden2013assessing} normalized   {\em Popt}  using:

{\footnotesize\begin{eqnarray} 
P_{opt}(m) = 1- \frac{S(optimal)-S(m)}{S(optimal)-S(worst)}\label{eq:popt}
\end{eqnarray}}
\noindent
where $S(optimal)$, $S(m)$ and $S(worst)$ represent the area of curve under the optimal learner, proposed learner, and worst learner.
Note that the worst model is built by sorting all the changes according to the actual defect density in ascending order.  
After normalization, {\em Popt(20)} (like   {\em d2h}) has the range zero to one.
Please note two important points. Firstly, unlike the defect prediction data of \tbl{versions}, the data for text mining task is 
not conveniently divided into versions. Hence, to generate separate train and test
data sets,  we use a $x*y$ cross-validation study where, $x=5$ times, we randomize the order of the data
then divide into $y=5$ bins. Then, we test on that bin after training on all the others.
Secondly:
\textcolor{\TEXT}{
\bi
\item
{\em larger} values of {\em Popt(20)} are {\em better}; 
\item
{\em smaller} values of {\em d2h} are {\em better}.
\ei
}
\subsubsection{Statistical Analysis}
\label{sect:sample}
As to statistical methods, the following results use two approaches.
Firstly, when comparing one result to a sample of $N$ others, we will sometime see ``small effects''  (which can be ignored). To define ``small effect'',
we  use Cohen's delta~\cite{cohen1988statistical}:

{\footnotesize \begin{equation}\label{eq:cohen}
d=\mathit{small\; effect} = 0.2*\sqrt{\frac{\sum_i^x(x_i- ({\sum}x_i/n))^2}{n-1}}\end{equation}}
i.e., 20\% of the standard deviation.

Secondly, other statistical tests are required
when   comparing  results from two samples; e.g., when
two variants of some stochastic process are applied, many
times, to a population.
For this second kind of comparison, we need a statistical 
significance test (to certify that the distributions are indeed different) and an effect size test
(to check that the differences are more than a ``small effect''). 
There
 are many ways to implement second kind of test.
Here,  we used those which have been past peer reviewed in   the literature~\cite{agrawal2018better,agrawal2018wrong}. Specifically, we use   Efron's 95\% confidence bootstrap procedure~\cite{efron93bootstrap} and the A12     test~\cite{Arcuri:2011}.  
In this second test, to say that 
one sample $S_1$ is ``{\em worse}'' than another sample $S_2$
is to say: 1) The mean {\em Popt(20)}  values of $S_1$  are  less than  $S_2$; 2) The mean {\em D2h} values of $S_1$  are more than  $S_2$; and 3) The populations are not statistically similar; i.e., (a)~their mean difference is larger than a small   effect (using A12) 
and that 
(b)~a statistical significance test (bootstrapping) has not rejected the hypothesis that
they  are different   (at 95\% confidence).
Note we do not use  A12 or bootstrap
for the first kind of test, since those statistics
are  not defined for comparisons of
individuals to a sample.

\section{Motivation for new work: Surprising Results from FFtrees}\label{sect:fft}

 This section  describes the    FFtrees results published by Chen et al. in FSE'18~\cite{chen2018applications} that (a)~motivated this article and (b)~lead to  our hypothesis that 
 ``redundant parameter choices might be leading to indistinguishable results''. 
 This will in turn lead to (next section)
 a new method called {\IT*} that deprioritizes choices that lead to redundant results.
 
Fast and Frugal Trees (FFtrees)
 were developed by  psychological scientists~\cite{martignon2008categorization}  
trying 
 to generate succinct, easily comprehensible models. 
FFtrees are  binary trees that 
return a binary classification (e.g., true, false).
Unlike standard decision trees, 
each level of an FFtree
must have at least one leaf node.
 For example,  
 \tbl{three} shows an FFTtree generated from
        the log4j JAVA system of \tbl{versions}.
        The goal of this tree is to classify a software module as ``defective=true'' or ``defective=false''.
         The four nodes in the \tbl{three} FFTree  reference  four  attributes \emph{cbo,\ rfc,\ dam,\ amc}  (defined in  
        \tbl{ck}).

  \begin{table}[!t]

\caption{An example FFtree  generated from  \tbl{versions} data sets. Attributes come from \tbl{ck}.
``True'' means ``predicted to be defective''.}\label{tbl:three}
 {
\begin{verbatim}
        if      cbo <= 4    then false      
        else if rfc > 32    then true         
        else if dam >  0    then true      
        else if amc < 32.25 then true 
        else false
        \end{verbatim}}
\end{table}

Following the advice of~\cite{phillips2017FFtrees,chen2018applications}, we generate   trees of   depth of   $d=4$.
This means that
 FFtrees make their decisions using at most four attributes (where numeric ranges have been binarized by splitting at the median point). 

Standard rule learners     select
ranges that best select for some goal (e.g., selecting for the ``true'' examples). This can lead to overfitting.
To avoid overfitting, FFtrees use a somewhat unique strategy:
 at each level of the tree, 
FFtrees builds two trees using the ranges  that  {\em most} and {\em least}  satisfy some goal; e.g., {\em d2h} or {\em Popt20}.
That is, half the time, FFtrees will try to avoid the target class by building a leaf node that exits to ``false''.
Assuming a maximum tree depth of  \mbox{$d=4$} and two choices at each level, 
then FFtree builds   $2^d=16$ trees then prunes away all but one, as follows:
\bi
\item
Firstly, select
a goal predicate; e.g., {\em d2h} or {\em Popt20}.
\item
Next, while {\em building one tree}, at each level of the tree,   FFtree scores
each range according to how well  that range \{does, does not\} satisfy that goal.
These  selected range becomes a leaf note. FFtree then calls itself
recursively on all examples that do not fall into   that range.
\item 
Finally, while {\em assessing 16 trees}, the training data is run through each tree to find what examples
are selected   by that tree. Each tree is scored by passing the selected examples  through the goal predicate.
\item
The tree with the best score is   applied to the test data.
\ei
In summary, FFtrees {\em explore around} a few dozen times, trying  different options for how to best model the data (i.e., what exit
node to use at each level of the tree).
After a few  {\em explorations}, FFtrees deletes the worst models, and uses the remaining model on the test data. 
 \begin{figure}[!t]
{\scriptsize
\begin{tabular}{cc}

{\em D2h}: {\em less} is {\em better}. 
&
{\em Popt(20)}: {\em more} is {\em better}. \\
``small effect'' $= 5.1$
&
``small effect'' $=5.2$\\
\includegraphics[width=1.6in]{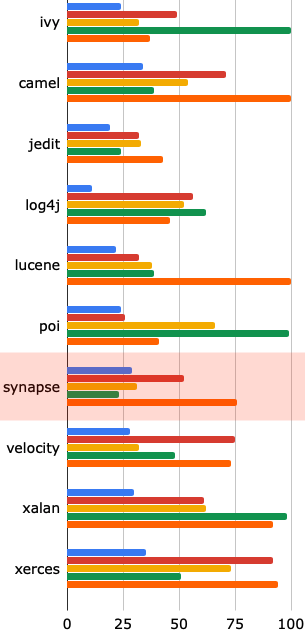}&\includegraphics[width=1.6in]{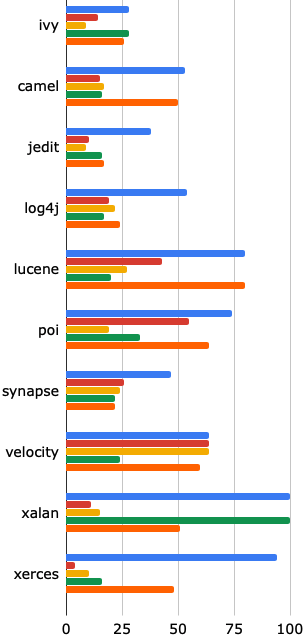}\\
\multicolumn{2}{c}{ \includegraphics[width=2in]{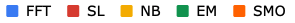}}
\end{tabular}}
\captionsetup{font=footnotesize}

 \caption{Defect prediction results for FFtree vs  untuned learners. From~\cite{chen2018applications}.
FFtrees is  almost never beaten  by  other methods  (by more than a ``small effect'').
Exception: see the synapse+EM results in the left column.
}\label{fig:chen}
\end{figure}

\fig{chen} shows results from Chen et al.~\cite{chen2018applications}  that compared
FFtrees to  standard defect predictors. 
In that comparison,  Ghotra et al.~\cite{ghotra2015revisiting} was used to guide learner selection.
They  found
that 32 defect predictors  group together into just four ranks from best to worse. (Please look for all four groups in Table 9 of \cite{ghotra2015revisiting}). 
We picked at random from each of their  ranks to select SL=Simple Logistic, NB=Naive Bayes, EM=Expectation Maximization, 
SMO=Sequential Minimal Optimization (a kind of support vector machine).  We call these learners ``standard'' since, in \fig{chen}, we use them
with their  defaults from   Scikit-learn~\cite{pedregosa2011scikit}.
In \fig{chen}:
\bi
\item Performance is evaluated  using metrics  from \tion{goals}.
\item  Data comes from \tbl{versions}.
\item This data has the attributes of \tbl{ck}.
 \item  For data with multiple versions, we test on the latest version and train on a combination of all the rest.
\item
If FFtrees perform worse than any other learner by more than a ``small effect''
  (defined using \eq{cohen}), then that result is highlighted in red (see  the synapse d2h results of \fig{chen}).
  For each column,  the size of a ``small effect'' is listed at top.
\ei
As shown in \fig{chen}, FFtrees  nearly always performs  as well, or better, than anything else.


\section{Research Method: The {\IT*} Algorithm}\label{sect:dodge}
 
It is very surprising that  something as simple as FFtree perform so well (see
\fig{chen}), especially since:
\bi
\item
FFtrees   explores very few  alternate
models (only 16).
\item
Each model 
references  only four attributes.
\item
To handle numeric variables,
a very basic discretization policy is applied at each level of tree 
(numerics are separated at the median value).
\item
Strange to say, half the time,  FFtree's overfitting mechanism will  try to {\em avoid} the target class
when it selects a leaf node that exits to ``false''.
\ei
 Under what conditions would something
that simple work as well as the other methods shown in \fig{chen}?
One possible answer was offered in the introduction. If the data
has large $\mathcal{E}$
  in its output space, then: 
 \bi
 \item
 The output/objective space  has  just a few cells; so
 \item  If there are $c$ cells and $t$ tunings, and when $t>c$, then some of those will be {\em redundant};
 i.e., they achieve results within $\mathcal{E}$ of other results.
 \item Which means that {\em exploring around}  $c$ times will cover much of the output space.
 \ei 
If that is true, then to do better than FFtrees:
\bi 
\item Try {\em exploring around} across a {\em wider range of options}. 
\item  If  some options result in a performance score $\alpha$, then we will {\em deprecate
 options} that lead to  $\alpha \pm\mathcal{E}$.
\ei
To find  a {\em wider range of options},
{\IT*} uses  the  \tbl{options} tree of options. Leaves in that tree are either:
\bi
\item Single choices; e.g.,  {\em DecisionTree}, ``splitter=random''; or
\item Numeric ranges; e.g., {\em Normalizer}, ``norm=l2''.
\ei

Each node in the tree is assigned a weight $w=0$.
When  {\em evaluating} a branch,
the options in that branch   
configure, then executes,  a pre-processor/learner. 
Each evaluation selects one leaf from the learner sub-tree and  
one from the pre-processing tree (and defect prediction and text mining
explores different pre-processing sub-trees, see \tbl{options}).
If the evaluation score
is more than $\mathcal{E}$ of prior scores, then all nodes in that branch are endorsed ($w=w+1$).
Otherwise, {\IT*} deprecates ($w=w-1$). 
{\IT*}  uses these  weights  to select options via a recursive {\em weighted descent}   where, at each level,
it selects sub-trees whose root has the largest weight (i.e., those    most
endorsed). 

The design conjecture
 of {\IT*} is that exploring some tuning options matters but, given
 a large $\mathcal{E}$ output space, the details of those
 options are not so important. Hence, a limited number of
 $N_1$ times, we pick some options at random. Having selected
 those options, for further $N_2$ samples, we learn
 which of the $N_1$ options should be most deprecated
 or endorsed.  

When a parameter range is initially  evaluated, a random number \mbox{$r=random(\mathit{lo}, \mathit{hi})$} is selected and its weight $w(r)$ is set to zero.
Subsequently, this 
weight is  endorsed/deprecated technique  as described above, with
one refinement. When  a new value  is required
(i.e., when the branch is evaluated again) then {\IT*} restricts
the $\{\mathit{lo}, \mathit{hi}\}$ range as follows.
If the  best,worst weights seen so far (in this range)
are associated with $b,w$ (respectively) then use
$\mathit{lo}=b$ and $\mathit{hi}=\frac{b+w}{2}$.
Important point: {\em endorse} and {\em deprecate} is done each time a branch is evaluated
within each   $N_1$ and $N_2$ steps. Figure~\ref{fig:tabu_pseudo} provides a summarized procedure on how \IT* works.

In summary, {\IT*} is a method for learning what tunings are {\em redundant}; i.e., lead to results that are very similar to other tunings~\cite{agrawal2019nature}.
It is controlled by two meta-parameters:
\bi
\item $\mathcal{E}$: results are ``similar'' if they differ by less than   $\mathcal{E}$;
\item $N$: the number of sampled tunings.
\ei
Recall that  $N=N_1+N_2$ where
\bi
\item
The first $N_1$ times, the set of tuning options grows;
\item
For the remaining $N_2$ times, that set is frozen while we refine our understanding of what tunings to avoid.
\ei

\begin{figure}[!tbp]
\captionsetup{justification=centering}
\small \begin{tabular}{|p{.95\linewidth}|}\hline
INPUT: 
\bi
\item A dataset
\item $\mathcal{E} \in \{0.05, 0.1, 0.2\}$
\item A goal predicate $p$; e.g., ${P_{\mathit{opt}}}$ or  $\mathit{d2h}$;
\item Objective, either to maximize or minimize $p$.
\ei
OUTPUT:
\bi\item Optimal choices of preprocessor and learner with   corresponding parameter settings.
\ei 
PROCEDURE:
\bi
\item Separate the data into train and test
\item Choose set of preprocessors, data miners with different parameter settings from Table~\ref{tbl:options}.
\item Build a tree of options for preprocessing and learning.   Initialize all   nodes with a weight of 0.
\item Sample at random from the tree to create  random combinations   of preprocessors and learners. 
\item Evaluate $N_1$ (in our case $N_1=12$) random samples  on training set and reweigh the choices as follows:
    \bi
    \item   Deprecate ($w=w-1$)  those options that result in the similar region of the performance score $\alpha$ ($\alpha \pm\mathcal{E}$)
    \item Otherwise endorse those choices ($w=w+1$)
    \ei
        
\item Now, for $N_2$ ($ N_2 \in \{30, 100, 1000\}$) evaluations 
    \bi
    \item Pick the learner and preprocessor choices with the  highest weight and mutate its parameter settings. Mutation is done, using some basic rules, for numeric ranges of attribute (look for a random value between $(best, (best+worst)/2)$ seen so far in  $N_1+ N_2$). For categorical values, we look for the highest  weight.
    \ei
\item For $N_1+ N_2$ evaluations,  track     optimal settings (those that lead to  best results on training data).
\item Return the optimal setting and apply these to test data.
\ei

\\\hline
\end{tabular}
\caption{Pseudocode of \IT*}
\label{fig:tabu_pseudo}
\end{figure}

\begin{figure}[!htbp]
\centering
\includegraphics[width=\linewidth]{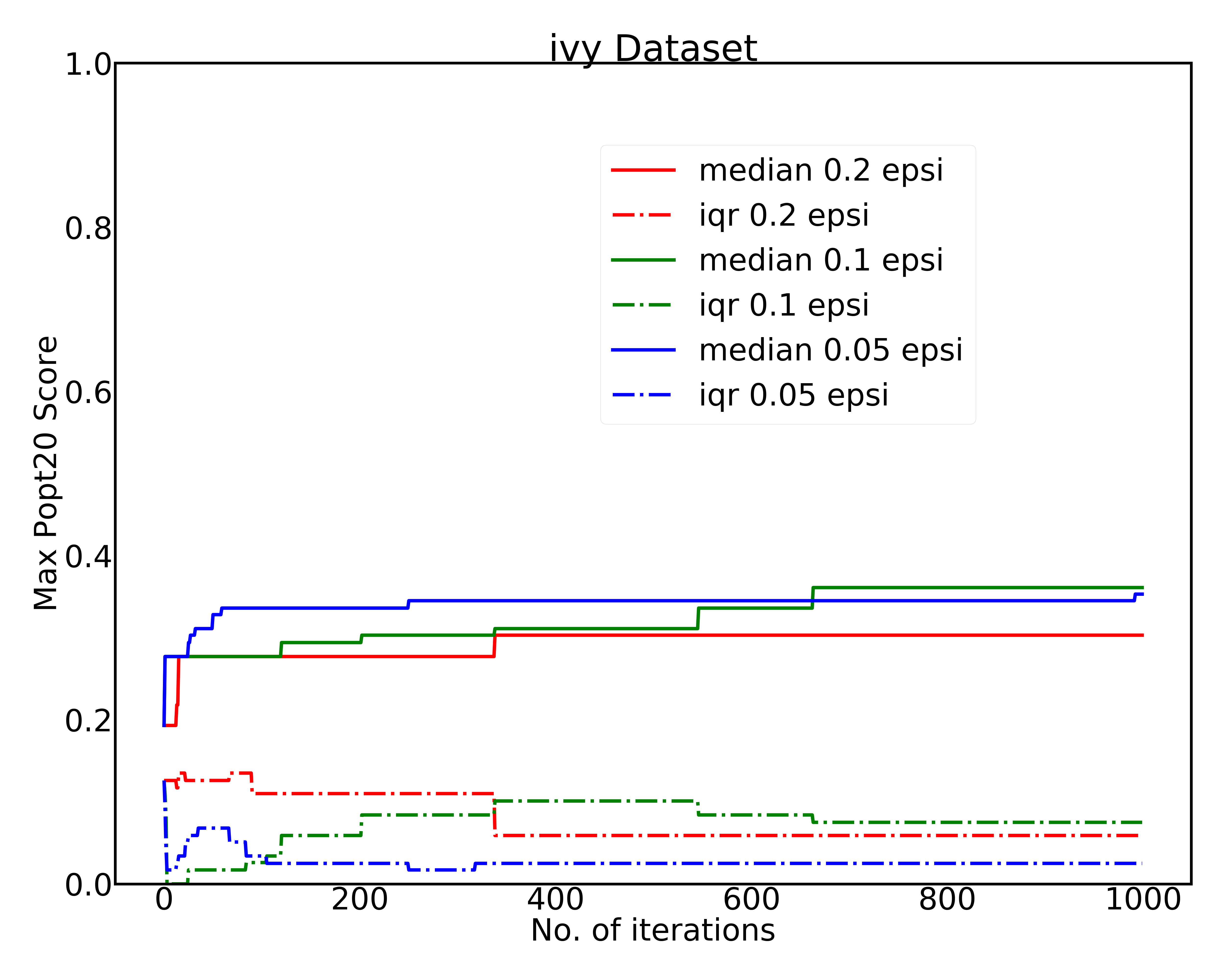}
\caption{\IT* for $P_{opt}$ on ivy dataset
(for results
on other datasets,
\href{http://tiny.cc/rq1a_tabu}{http://tiny.cc/rq1a\_tabu}).
Here the  X-axis represents number of samples needed
and the Y-axis represents the Max value of $P_{opt}$ seen until that sample. On the y-axis, {\em larger} values are {\em better}.   Note that the performance seen after 50 samples is nearly the same as seen after 500 or 1000 samples. }
    \label{fig:motivate_quickly}
\end{figure}

\begin{figure}[!t]
{\scriptsize
\begin{tabular}{cc}

{\em D2h}: {\em less} is {\em better}.   
&
{\em Popt(20)}: {\em more} is {\em better}.  \\

``small effect'' $=4.4$
&
``small effect'' $=6.1$\\

\includegraphics[width=1.6in]{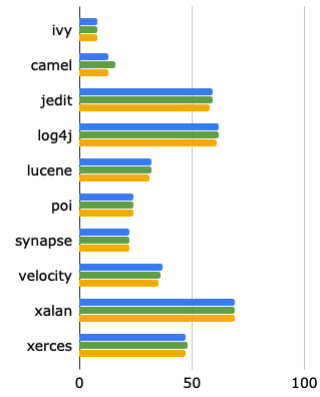}&\includegraphics[width=1.6in]{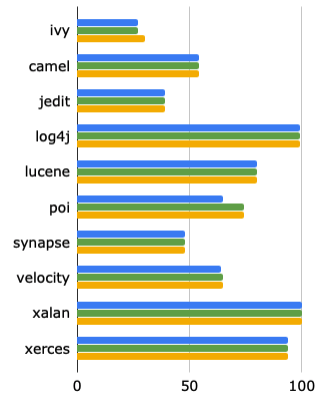}\\
\multicolumn{2}{c}{ \includegraphics[width=1.3in]{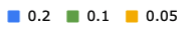}}
\end{tabular}}

 \caption{{\bf RQ1} results. Defect prediction with \IT{$\mathcal{E} \in \{0.2, 0.1, 0.05\}$},  terminating \IT* at $N=30$ evaluations.  
As before,   changing  $\mathcal{E}$  does not change learner performance any more than a ``small effect''.
   This figure was generated using the 
 same experimental set up as \fig{varN}.}
 \label{fig:varE}
\end{figure}

\begin{figure}[!t]
{\scriptsize
\begin{tabular}{cc}

{\em D2h}: {\em less} is {\em better}. 
&
{\em Popt(20)}: {\em more} is {\em better}.   \\

``small effect'' $=4.0$
&
``small effect''  $=4.9$\\

\includegraphics[width=1.6in,height=2.1in]{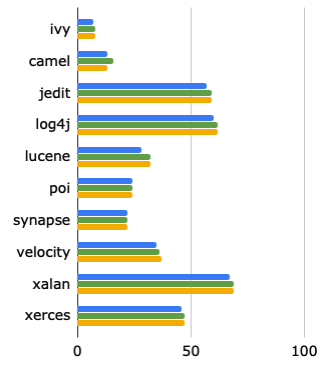}&\includegraphics[width=1.6in,height=2.1in]{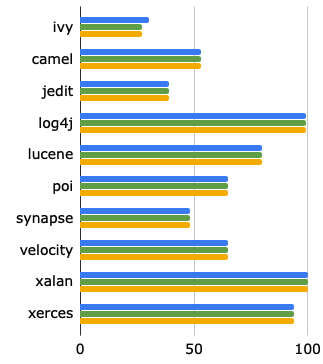}\\
\multicolumn{2}{c}{ \includegraphics[width=1.7in]{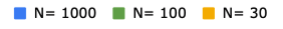}}
\end{tabular}}

 \caption{More {\bf RQ1} results. Defect prediction  with {\IT{.2}}, varying samples $N$.
  Note that for any data set, all these results are very similar; i.e., changing the number of evaluations  does not change learner performance any more than a ``small effect''.
  This figure was generated using the 
 same experimental set up as \fig{chen} (where tuning options  taken from \tbl{options}).
 }\label{fig:varN}
\end{figure}

\section{Experimental Results}\label{sect:rq}
Using {\IT*}, we can now answer the research questions asked in this article's introduction.

\subsection{RQ1:  Is \IT* too complicated?   How to find appropriate value of $\mathcal{E}$?}\label{sect:rq1}

\textcolor{\TEXT}{
Firstly, we wanted to verify whether our hypothesis of "redundant options (similar region defined within $\mathcal{E}$) might be leading to indistinguishable results". To test this,
we use
Figure~\ref{fig:motivate_quickly} to see how quickly 
(i.e., after how many
evaluations $N$)
the 
performance of
\IT* plateaus.
In figure~\ref{fig:motivate_quickly}, the X-axis represents number of samples ($N$)
and the Y-axis represents the max value of $P_{opt}$ seen until that sample
(and for that measure,
 {\em larger} values are {\em better}).  We used $\mathcal{E}$ of  0.05, 0.1, and 0.2 values, and looked
for the number of samples needed before the performance plateaus.  We also show the performance
variability measured in terms
of the interquartile
range (IQR) (which is 
the (75-25)th percentile). 
These IQR values
are   very small; i.e.,
 \IT*'s performance is very stable.}

\textcolor{\TEXT}{In Figure~\ref{fig:motivate_quickly},
we observe
that most change in improvement
happens after just few
tens of evaluations.
This supports our  hypothesis that there are "redundant options which lead to indistinguishable results".}
\noindent
\fig{varE} and \fig{varN}  explore different settings
of $\{N,\mathcal{E}\}$.

\textcolor{\TEXT}{
\bi
\item \fig{varE} varies  $\mathcal{E}$   but keeps $N$ constant.
In this treatment,
we check   how much improvement do we miss on when we try to find the right $\mathcal{E}$ value. Let's say we cut the Figure~\ref{fig:motivate_quickly} at $N=30$ line on x-axis, we report the values of  $\mathcal{E}$ for 0.05, 0.1 and 0.2.
\item \fig{varN} varies $N$ but keeps $\mathcal{E}$ constant. Please note, when we say $\mathcal{E}$ constant, we wanted to see how much improvement do we miss on when we try to find the right $N$ value. Let's say we look at $\mathcal{E}=0.2$ line, we report the values of  $N$ for 30, 100 and 1000.
\ei
As shown in these figures, changes to $\{N,\mathcal{E}\}$ alter the performance of \IT* by less than a ``small effect''.}

That is,
(a)~the output space for this data falls into a very small number of regions so
(b)~a large number of samples across a  fine-grained division of the output space 
performs just as well as a few samples over a coarse-grained division.

 In summary, our answer to {\bf RQ1} is that the values of $\{N,\mathcal{E}\}$ can be set very easily. Based on the results of \fig{varE} and \fig{varN}, for the rest of this article
we will use $\mathcal{E}=0.2$ while taking $N=30$ samples of the options from Table~\ref{tbl:options}. We observed that there is no significant loss in performance if we move $\mathcal{E}$ to different values or $N$ provided in figures~\ref{fig:varE} and~\ref{fig:varN}. We picked $\mathcal{E}=0.2$ as we are looking for larger redundant region in data at the same time faster evaluations of $N=30$.





\subsection{RQ2: How does {\IT*} compare to recent prominent  defect  prediction  and  hyperparamter  optimization results?}

SMOTUNED  is 
Agrawal et al. ICSE'18~\cite{agrawal2018better}'s hyperparamater optimizer that tunes SMOTE, a data pre-processor
(recall that 
SMOTE is a tool for addressing class imbalance and was described in \tion{daa}).
Agrawal et al. reported that SMOTUNED's tunings
  greatly improved classifier
performance. SMOTUNED uses differential evolutionary  algorithm~\cite{storn1997differential} and tunes the 
control parameters of SMOTE (see \tbl{options}).

DE+RF is a hyperparameter optimizer proposed by Fu et al.~\cite{fu2016differential}
that  uses differential evolution to tune the control parameters of random forests.
The premise of RF (which is short for random forests) is ``if one tree is useful, why not a hundred?''. 
RF quickly builds many trees, each time using a random selection of the attributes and examples. The final conclusion  is then generated by polling across all the trees in the forest. RF's control parameters are listed in \tbl{options}.

SMOTUNED and 
DE+RF used DE since
(a)~DE  can handle numeric and discrete options; and
(b)~it has  proven useful in prior SE 
studies~\cite{fu2016tuning}. 
Further,
other evolutionary algorithms (genetic algorithms~\cite{goldberg2006genetic}, simulated annealing~\cite{kirkpatrick1983optimization}) mutate each attribute in isolation.  When two attributes are correlated, those algorithms can mutate variables inappropriately in different directions. DE, on the other hand, mutates  attributes
in tandem along known data trends.    
Hence, DE's tandem search can outperform other optimizers such as  
(a)~particle swarm optimization~\cite{vesterstrom2004comparative}; (b)~the grid search used by Tantithamthavorn et al.
to tune their defect predictors~\cite{tantithamthavorn2016automated,8263202}; or (c)~the genetic algorithm 
used by Panichella et al. ~\cite{Panichella:2013} to tune a text miner (see below).

\begin{figure}[!t]
{\scriptsize
\begin{tabular}{cc}

\textcolor{\TEXT}{{\em D2h}: {\em less} is {\em better}. }
&
\textcolor{\TEXT}{{\em Popt(20)}: {\em more} is {\em better}.}   \\

Mean results from 25 runs
&
Mean results from 25 runs  \\

\includegraphics[width=1.6in]{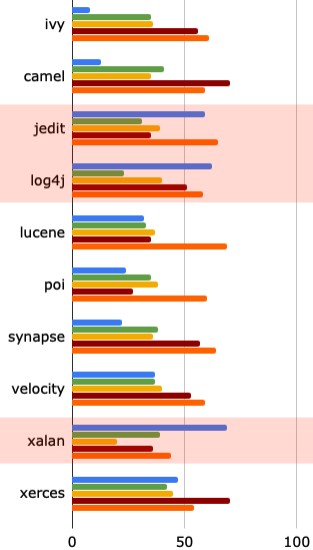}&\includegraphics[height=2.8in,width=1.6in]{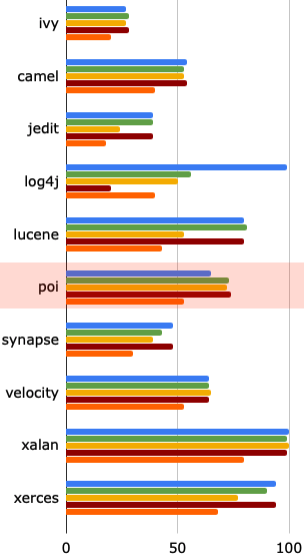}\\
\multicolumn{2}{c}{ \includegraphics[width=2.5in]{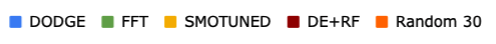}}
\end{tabular}}

 \caption{{\bf RQ2} results. Defect prediction results for {\IT{.2}, $N=30$} vs  (FFtrees, SMOTUNED, DE+RF, RANDOM).
 In only a few cases (those highlighted in red)  is {\IT{.2}}'s performance worse than anything else (where ``worse''
 is defined using the statistics of \tion{sample}.)  }\label{fig:smote}
\end{figure}

\fig{smote} compares  hyperparameter optimizers with {\IT{.2}}, FFtrees and (just for completeness) a random search method that picks   30 random  options (equivalent N as of \IT*) from \tbl{options}. These experiments make extensive use of stochastic algorithms 
whose behavior can significantly differ between each run (DE and Random30).
Hence, \fig{smote} shows mean results from 25 runs using 25 different seeds. In those results:
\bi
\item
Usually, random performs badly and never defeats {\IT*}. This result tells us that the reweighing  scheme within {\IT*} is useful.
\item
In 16/20 cases combining the \textit{d2h} and \textit{Popt20} datasets,   {\IT{.2}} is no worse than anything else
(where ``worse'' is defined as per \tion{sample}).
\item
In two cases,  {\IT{.2}} is beaten by FFtrees (see the {\em d2h} results for jedit and log4j).
That is, in 90\% of these results, methods that explore a little around the results space do no worse than methods that try to
extensively explore the space of tuning options.
\ei
 In summary, our answer to {\bf RQ2} is that {\IT*} often performs
 much better than  recent prominent standard    hyperparameter  optimization results.

\begin{figure}[!t]

\begin{center} 
{\footnotesize \textcolor{\TEXT}{{\em D2h}: {\em less} is {\em better}}. 

Mean results from 25 runs.

\vspace{1mm}
}

\includegraphics[width=2.4in]{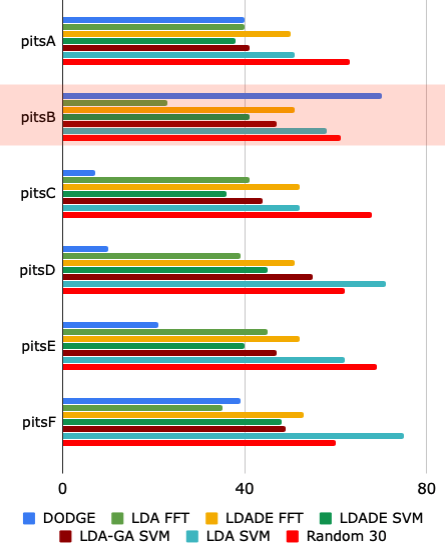}
\end{center}
 
 \caption{{\bf RQ3} results.  Mean text mining  prediction results using {\IT{.2}} and $N=30$.
 In only one case (PitsB), \IT*'s performance is worse than anything else
 (where ``worse'' is defined as per \tion{sample}).
 Same experimental set up as \fig{chen} except here, 
 we use Efron's 95\% confidence  bootstrap procedure~\cite{efron93bootstrap} (to demonstrate
 significant differences), then the   A12 effect size test~\cite{Arcuri:2011} (to demonstrate
 that the observed delta is bigger than a ``small effect'').
 }\label{fig:text}
\end{figure}

\subsection{RQ3: Is {\IT*} only useful for defect prediction?}
\label{sect:text}

{\IT*} was designed in the context  of defect prediction. This section checks if that design   applies to a very
different software analytics; i.e.,  SE text mining.
Note that, as with defect prediction, hyperparameter optimizers (like {\IT*}) adjust
the control parameters of Table~1. 
\textcolor{\TEXT}{In the particular case of text mining,
 we adjust the Table~1 text mining data pre-processing options (used to generate data sets like
\tbl{features}). We also adjust the Table~1 learner options.}

\fig{text} shows our text mining results.  As before, for completeness sake,
 we include results by RANDOMly selecting tuning and learning options.


 
As seen in \fig{text}, in only one case
  {\IT*}'s performance is worse than anything else
 (where ``worse'' is defined as per \tion{sample}).
 The LDA-FFT results for PitsF is 2\% better than {\IT*}, but
 difference was deemed insignificant by our statistical tests.
And, just as with the \fig{smote} results, when {\IT*} fails, it is beaten by a treatment that uses FFtree (see the PitsB LDA-FFT results).
 That  is,  in  100\%  of  these  results,  methods  that explore  a  little  around  the  results  space  do  no  worse than methods that try to extensively explore the space of tuning options (e.g., genetic algorithms and differential evolution).
 
 In summary, our answer to {\bf RQ3} is that {\IT*} is not just a defect prediction method. Its success with text mining make it an interesting
 candidate for further experimentation with other SE tasks.

\section{Threats to Validity}
\label{sect:threats}

\textcolor{\TEXT}{This paper is not about how to generate good predictors, per se. Instead, it is more about  an instrument ({\IT*}) that probes the nature of the 
space options associated with  AI tools applied to SE. We show that when prior work has tried to generate good predictors, their algorithms have been much slower
than necessary since they waste much time exploring a large number  of redundant options.}

\textcolor{\TEXT}{Nevertheless, our experimental rig  repeats numerous prior studies (this time adding in {\IT*}).  
Such is the nature of repeated studies that our
 work shares the same threats to validity as that of prior work (discussed below).}

\textcolor{\TEXT}{\textbf{Sampling Bias:}
    This article shares the same sampling bias problem as every other data mining paper. Sampling bias threatens any classification experiment (since what matters in one data set may or may not hold in another).  For example, one of our sampling biases is that all our data comes
    from open source projects.} 
    
\textcolor{\TEXT}{Having said that,  our sample bias is somewhat smaller than other papers since, we applied our frameworks to 16 SE data sets giving us more conclusive results. 
Also, we showed that there exists more than one
    domain where {\IT*} is a useful approach.}

\textcolor{\TEXT}{\textbf{Learner Bias:}
For building different classifiers in this study, we used many preprocessors (13) and learners (6). We chose these learners because past studies have shown that, these have been extensively used~\cite{ghotra2015revisiting,agrawal2018better,tantithamthavorn2016automated,8263202}. Thus they are selected as the state-of-the-art learners to be compared with \IT*. 
In theory, there exists
other learners (which we have not explored) and could change our results.}

One important class of learners not studied 
here are those that use numerous
hyperparameters. 
All the learners explored here by DODGE  have less than
dozen hyperparameters.
In the future, it would be worth  
studying the value of DODGE on more complex  machine  learning  algorithms  such as  neural  networks.

\textcolor{\TEXT}{\textbf{Evaluation Bias:}
This paper uses two performance measures, i.e., $P_{opt}$ and $dist2heaven$. Other quality measures are often
used in software engineering to quantify the effectiveness of prediction ~\cite{Menzies:2007prec, menzies2005simple, jorgensen2004realism}. 
We used these measures since we wanted to show the success of \IT* for multi-goals and these two measures are more prominent in the literature.}

\textcolor{\TEXT}{\textbf{Order Bias:} For the performance evaluation part, the order that
    the data trained and predicted can affect the results.
    Also, 
    for the defect prediction datasets,
    we deliberately choose an ordering that mimics
    how our software projects releases versions so, for those experiments, we would say that bias was   required and needed.
    Further, 
    for the other text mining datasets, to 
      mitigate this order bias, we ran our rig in a   5-bin cross validation 5 times, randomly changing the order of the data each time.}

\textcolor{\TEXT}{\textbf{Construct Validity:} At various stages of data collection by different researchers, they must have made engineering decisions about what object-oriented metrics need to be extracted. Though all
these decisions have been verified and evaluated by past researchers~\cite{agrawal2018better,agrawal2018wrong} to make sure the dataset collection do not suffer from any construct validity.}



\textbf{External Validity:}
{\IT*} self-selects  
the tunings used in the  pre-processors and data miners. Hence, by its very nature, this
article avoids one threat to external validity
(i.e.,   important control parameter settings are explored). 

This paper 
reports results from two tasks (defect prediction and text mining)
to show that the same effect holds in both tasks; i.e., 
algorithms can be remarkably effective
when they assume  the output space seems to
divide into a very small number of regions. 
Most software analytics papers report results from one task;
i.e., either defect prediction {\em or} text mining. In that sense, the external validity of this paper is greater than most analytics papers.

On the other hand, this paper {\em only} reports results
from two tasks.   There
are many more kinds of SE tasks that should be explored
before it can be conclusively stated that {\IT*} is widely
applicable and useful.

Another threat to external validity is that
this article compares {\IT*} against 
existing  hyperparameter optimization in the software analytics literature. We do not compare our new approach
against the kinds of optimizers we might find
in search-based SE literature~\cite{petke2018guest}.
There are two reasons for this.
Firstly, search-based SE methods are typically CPU intensive and so do not address our faster termination goal.
Secondly,   the main point of this article is
to document  a previous unobserved  feature of the output
space of software analytics. 
In order
to motivate the community to explore that space, some article
must demonstrate its existence and offer an initial results showing that, using
the knowledge of output space, it is possible to do better than past work.

 \section{Related Work}\label{sect:related}


\begin{redish}
{\IT*} is a novel
hyperparameter optimizer. This section offers some brief notes on other research into hyperparameter optimizers.
Note that applications of hyperparameter optimization to software engineering is a very
large topic. 
Elsewhere~\cite{Agrawal19z} we offer an extensive literature review on hyperparameter
optimization and its applications in software engineering. Here, we offer some overview notes.

Apart from {\IT*}, there  are  many ways to implement hyperparameter optimizers.
For example, \textit{grid search}~\cite{bergstra2012random} 
   creates   
$C$ nested for-loops to explore   $C$ control parameters.  Bergstra et al. deprecate grid search arguing that (a)~the best hyperparameters are usually found within a very small region of the total space;  and (b)~a grid search that is fine-grained enough to find that region for any learner and any data set would be very slow indeed~\cite{bergstra2012random}.
Despite this, some SE researchers persist in using grid search~\cite{tantithamthavorn2016automated,8263202}.

Another way to implement hyperparamter optimization is \textit{random search}~\cite{bergstra2012random}.
This approach sets up ranges  of hyperparameter values and select random combinations to train the model and evaluate. 
There are many other ways to implement this kind of optimization
  including those that use some form of genetic algorithm
like differential evolution~\cite{storn1997differential},
NSGA-2~\cite{deb00afast}, IBEA~\cite{Zitzler04indicator-basedselection}, or MOEA/D~\cite{zhang07}.
In this paper, we have already seen examples of the these standard hyperparameter optimizers.
For example, LDA-GA SVM used its own genetic algorithm while
LDADE FFT and LDADE SVM both used differential evolution~\cite{storn1997differential}.  
As shown in \fig{timings}, those algorithms took (much) longer to execute and (measured in terms of {\em d2h}, usually perform worse than 
as {\IT*}).  

The slowness of standard hyperparameter optimizers  restricts the space of hyperparameters that can be explored.
For example, Arcuri \& Fraser~\cite{Arcuri2013} warn that ``the possible number of parameter combinations is exponential in the number of parameters, so only a limited number of parameters and values could be used.''.
We conjecture that if they used {\IT*}, then they could have explored more parameters and possibly reversed
their conclusion that hyperparameter optimization adds little extra value.   Since
  Arcuri \& Fraser's
2013 study, other researchers in that research sub-area (test case generation) have found hyperparameter tuning
very useful; e.g., see the 2015 study by Panichella et al.~\cite{7102604}. 

The Panichella et al. study is 
very relevant to this paper since their 12,800 experiments (each with a give-up time of 600 seconds),
required 12.7 weeks of CPU to terminate. 
We conjecture that with tools like {\IT*},
more  studies like Panichella et al.
could be completed, much quicker,  with far fewer resources.

More generally, {\IT*} could speed up standard
hyperparameter optimization. We conjecture that  those optimizers could run much faster
if they pruned away redundant evaluations using {\IT*}.
If that were true then   {\IT*} could have a very large impact over a very wide range of research.

There is another way that  {\IT*}   comments  on  standard   optimization methods.
{\em Landscape analysis} is the process of exploring a large complex problem/solution space
in order to learn its shape. Once that is learned,  then different search strategies
could be proposed to better survey that particular shape. One drawback with landscape analysis
is that it can be extremely computationally expensive. To learn the landscape associated with the
test suites of 19 software programs,
30 times, 
Aleti et al.~\cite{Aleti2017}  evaluated 1,000,000 test suites.
The lesson of {\IT*} is that, sometimes, landscapes can be mapped without requiring 
19*30*1,000,000=570 million evaluations. For example, in this paper, we assumed a particular ``landscape'' (see Figure~1)
then designed a search method, {\IT*}, that would succeed quickly if that landscape existed,
or fail badly otherwise. Perhaps this strategy could be used in future research to reduce the cost
of landscape analysis.



\end{redish}

\section{Conclusion} 
This article has discussed ways to reduce the CPU cost associated with
hyperparameter optimization for software analytics. 
Tools like FFtrees or {\IT*} were shown to work as well, or better, than numerous
recent SE results:
\begin{redish}
\bi
\item
FFtrees work so well since  the output space looks like Figure~1 (i.e., it contains only a few regions
where results can be distinguished from each other).
In that space, FFtree's limited probing serves to sample the space. 
\item
{\IT*} works better than FFtrees since the  deprecation strategy of  \fig{tabu_pseudo}
is a better way to sample Figure~1 than FFtree's random probes.
\item
Other  methods  (used in prior SE research) perform   worse than {\IT*} since they do not appreciate the simplicity of
the output space (where ``simplicity'' means that it only contains a few
distinct results). Hence, those other methods waste much CPU as they
 struggle  to cover  billions of tuning options like \tbl{options} (most of which yield  indistinguishably   different  results).
 \ei
 \end{redish}

\noindent
Generalizing from our results,
perhaps it is time for a   new characterization of software analytics: 
\begin{quote}
{\em
Software analytics is that branch of machine learning
that studies problems  with large
$\mathcal{E}$ outputs. }
\end{quote}
This new characterization is  interesting since it means that a   machine learning algorithm developed in the AI 
community might not apply to SE. 
\textcolor{\TEXT}{A similar conclusion has recently been offered by  Binkley et al. who
argue for SE-specific   information retrieval methods~\cite{binkley2018need}.}

We suspect that understanding SE is a  different problem to understanding
other problems that are more precisely retrained.   Perhaps, it is time to design new
machine learning algorithms (like {\IT*}) that are better
suited to    large $\mathcal{E}$   SE problems.
As shown in this article, such new algorithms can exploit the peculiarities  of SE data to  dramatically improve   software analytics.

We hope that this article inspires much future work on 
a next generation of SE data miners. For example,
tools like {\IT*}  need to be applied to more SE tasks to check the external validity of these results. Another useful  extension to this work
would be to explore problems with three or more goals
(e.g., reduce false alarms while at the same time
improving precision and recall).
\textcolor{\TEXT}{Further, as discussed in the Related Work section, there are research opportunities where (a)~{\IT*} is used to  repeat and improve prior work
or  
(b)~speed up a wide range of other search-based SE algorithms (by using redundancy  pruning to reduce the space of candidate mutations).}

Lastly, there are many ways in which {\IT*} could be improved.  Right now we only deprecate tunings that lead to similar results. Another approach would be to deprecate
tunings that lead to similar {\em and worse} results (perhaps to rule out  parts of the output space, sooner).
Also, it would be useful if the \tbl{options} list could be reduced to a smaller, faster to run, set of learners. That is, here we could select learners which can terminate faster while generating the most variable kinds of models.


   \section*{Acknowledgements}
   This work was partially funded by an NSF  Grant \#1703487.

\bibliographystyle{IEEEtranS}
\newpage

\newpage

\vskip -2\baselineskip
\begin{IEEEbiography}[{\includegraphics[width=0.7in,keepaspectratio]{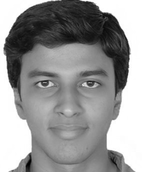}}]{Amritanshu Agrawal}
 holds a Ph.D. in Computer Science from North Carolina State University, Raleigh, NC. 
He explored better and faster hyperparameter optimizers for software analytics.  He works as a Data Scientist at Wayfair, Boston. For more, see
\url{http://www.amritanshu.us}.
\end{IEEEbiography}
\vskip -4\baselineskip 
\begin{IEEEbiography}[{\includegraphics[width=0.8in,clip,keepaspectratio]{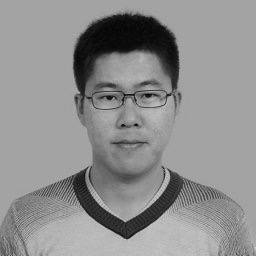}}]{Wei Fu} holds a Ph.D. from CS, NC State University. He now works at Landing.ai in Palo Alto.  \url{http://fuwei.us}
\end{IEEEbiography}
\vskip -4\baselineskip
\begin{IEEEbiography}[{\includegraphics[width=0.8in,clip,keepaspectratio]{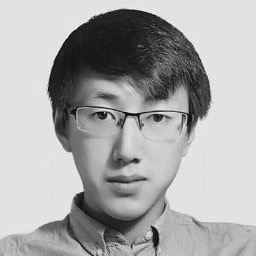}}]{Di Chen} holds a  master  in CS from NC State University where
he explored crowdsourcing and Machine learning. Mr Chen now works at
Facebook, California.
\end{IEEEbiography}
\vskip -4\baselineskip
\begin{IEEEbiography}[{\includegraphics[width=0.8in,clip,keepaspectratio]{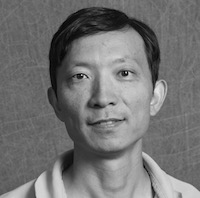}}]{Xipeng Shen}
is a Professor in CS at NC State.  His research interests are data mining, programming
languages and optimization. 
 Prof. Shen is an ACM Distinguished Member and a senior member of IEEE. 
 \url{https://people.engr.ncsu.edu/xshen5/}
\end{IEEEbiography}
\vskip -4\baselineskip
\begin{IEEEbiography}[{\includegraphics[width=0.8in,clip,keepaspectratio]{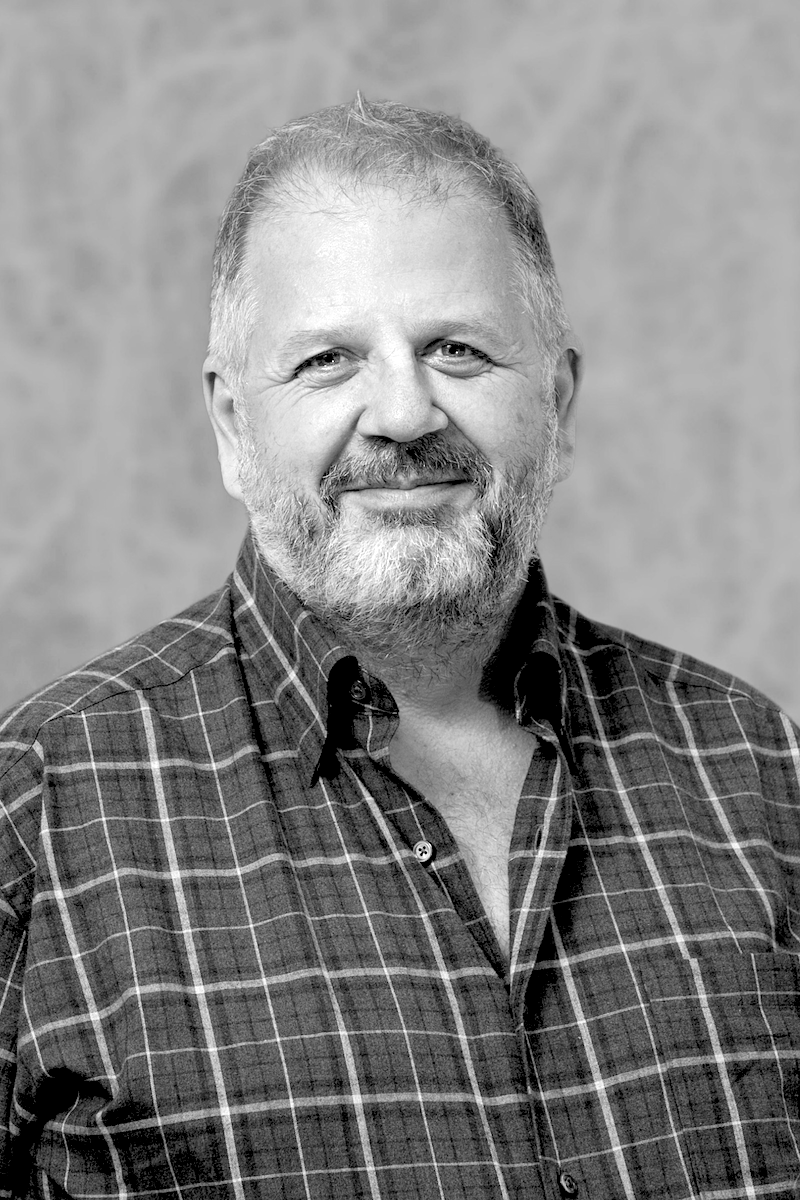}}]{Tim Menzies}
is a Professor in CS at NC State  His research interests include software engineering (SE), data mining, artificial intelligence, and search-based SE, open access science.
Prof. Menzies is an IEEE Fellow. \url{http://menzies.us}
\end{IEEEbiography}




\end{document}